\documentclass[a4paper,12pt]{article}

\usepackage{latexsym}

\newcommand\sqg{\sqrt{-g}}

\newcommand{\ud}{\mathrm{d}}

\begin{document}

\title{Restrictions on possible forms of classical
matter fields carrying no energy}

\author{Leszek M. SOKO\L{}OWSKI \\
Astronomical Observatory,Jagellonian University, Orla
171, \\ Krak\'ow 30-244, Poland \\sokolo@oa.uj.edu.pl} 

\date{}
\maketitle

\centerline{Short title: Classical matter fields without energy}

\begin{abstract}
It is postulated in general relativity that the matter 
energy--momentum tensor vanishes if and only if all the matter 
fields vanish. In classical Lagrangian field theory the energy and 
momentum density are described by the variational (symmetric) 
energy--momentum tensor (named the stress tensor) and a priori it 
might occur that for some systems the tensor is identically zero for 
all field configurations whereas evolution of the system is subject 
to deterministic Lagrange equations of motion. Such a system would 
not generate its own gravitational field. To check if these systems 
can exist in the framework of classical field theory  we find a 
relationship between the 
stress tensor and the Euler operator (i.e.~the Lagrange field 
equations). We prove that if a system of interacting scalar fields 
(the number of fields cannot exceed the spacetime dimension $d$) 
or a single vector field (in spacetimes with $d$ even) has the 
stress tensor such that its divergence is identically zero 
(i.e.~"on and off shell"), then the Lagrange equations of motion 
hold identically too. These systems have then no propagation 
equations at all and should be regarded as unphysical. Thus 
nontrivial field equations require the stress tensor be nontrivial 
too. This 
relationship between vanishing (of divergence) of the stress tensor 
and of the Euler operator breaks down if the number of fields is 
greater 
than $d$. We show on concrete examples that a system of $n>d$ 
interacting scalars or two interacting vector fields can have the 
stress tensor equal identically zero while their propagation 
equations are nontrivial. This means that 
non--self--gravitating (and yet detectable) field systems are in 
principle admissible. Their equations of 
motion are, however, in some sense degenerate. \\
We also show, that for a system of arbitrary number of 
interacting scalar fields or for a single vector field (in some 
specific spacetimes in the latter case), if the stress  
tensor is not identically zero, then it cannot vanish for all 
solutions. There do exist solutions with nonzero energy density and 
the system back--reacts on the spacetime. 

\end{abstract}


\section{Introduction} 
It is commonly accepted in general relativity
that the gravitational field is generated by all forms of matter
(i.e.~all species of elementary particles and fields) and that all
features of a given form of matter that are relevant for 
determining
its gravitational field are encoded in its energy--momentum 
tensor (cf.~e.g.~\cite{HE,S}). In other terms energy and linear 
momentum are the source of the gravitational field. This 
postulate does not tell one how
to construct the energy--momentum tensor for a given kind of 
matter and
whether it is unique. In principle this tensor should be 
determined in
a special--relativistic theory describing the material system and 
then
minimally coupled to gravity. However a reliable expression for 
this tensor has been found only in few cases (classical 
electrodynamics,
relativistic hydrodynamics etc.) and it has turned out that the
canonical energy--momentum tensor, though being conceptually 
important
due to the first Noether theorem, does not provide in most cases 
the correct
value of energy density and therefore is unphysical. A unique 
universal
definition of energy and momentum density arises if the equations 
of motion for the matter under consideration can be derived from a
Lagrangian. This is the Hilbert's variational (with respect to the
spacetime metric) energy--momentum tensor (or Belinfante tensor),
hereafter denoted as the stress tensor. It is worth emphasizing 
that the physical energy--momentum tensor cannot be determined by 
merely manipulating with the matter equations of motion and the 
use of the stress tensor is indispensable \cite{MS1}. Now it is 
commonly accepted
that it is this tensor that correctly describes energy and 
momentum
density and their flows for any Lagrangian matter both in curved 
and flat spacetimes \cite{We}. \\

It is postulated that the Lagrangian for any matter is such that 
the resulting stress tensor $T_{\mu\nu}$ vanishes on an open 
domain in the spacetime if and only if the matter fields vanish on 
the domain \cite{HE}. This condition expresses the principle that 
any matter carries energy. The energy conditions (\cite{HE}, 
Chap. 4) 
actually imposed on a matter Lagrangian exclude negative energies 
and ensure the "only if" condition. 

In this paper we investigate the problem whether there is 
some classical matter having no energy at all, i.e.~whether there 
exists a 
matter Lagrangian giving rise to the stress tensor vanishing 
identically, $T_{\mu\nu}\equiv 0$, for all field configurations 
independently of the field equations ("on and off shell"). Such 
a matter should be subject to deterministic equations of motion 
(the Lagrange ones) and according to Einstein field equations would 
propagate as a test one in a fixed spacetime. This matter would be 
non-self-gravitating and its interactions with other fields would 
be severely restricted by the constraint that they should exclude 
any energy transfer. At first sight one might conclude that the 
lack of any energy exchange with ordinary matter systems implies 
that this matter is nondetectable and as such it 
may be merely ignored. This is the case in quantum theory. However, 
here we are dealing with classical fields and in classical 
physics any system may be treated as open, i.e.~as being in 
contact with some surrounding. The contact may be arbitrarily weak, 
i.e.~involve negligibly small energy transfer (or no transfer at 
all), nevertheless by observing the surrounding (which acts as a 
"marker") one can make measurements on the system. In Example 1 
we give a hint of how a system of scalar fields without energy 
might be detected by their coupling to the electromagnetic field. 
We do not pursue the problem further, it is sufficient to say here 
that the lack of energy does not imply "physical nonexistence" in 
the sense of nondetectability. \\
This is why we investigate in this work to what extent 
Lagrangian field theory admits such bizarre systems and whether 
some of them can be excluded on theoretical grounds. To 
this end we show for some classical fields that there is a close 
connection between energy (the stress tensor) and deterministic 
equations of motion. For these fields we prove that the identically 
vanishing stress tensor implies that the Lagrange equations also 
identically vanish for all field configurations, i.e.~the fields 
are subject to no propagation equations at all. On very generic 
physical grounds one can then reject these fields as unphysical. \\

We first formulate the problem in full generality. Consider a 
system 
of classical matter tensor fields $\psi_A$ with a collective index 
$A$ (there is no need to deal with spinor fields since they can 
be expressed 
as tensor ones \cite{HE}). An example of a vanishing stress tensor 
is provided by a Lagrangian for $\psi_A$ of the form 
\begin{equation}\label{n1}
L(\psi_A, \psi_{A;\mu})=\frac{1}{\sqrt{-g}}L_0(\psi_A, \psi_
{A,\mu}),
\end{equation}
where $L_0$ is a scalar density of the weight $+1$ (so that $L$ 
is a 
genuine scalar) and $L_0$ is independent of the spacetime metric 
and its first derivatives,
\begin{equation}\label{n2}
\frac{\partial L_0}{\partial g_{\mu\nu}}=
\frac{\partial L_0}{\partial g_{\mu\nu,\alpha}}=0.
\end{equation}
(Here $\psi_{A;\mu}\equiv \nabla_{\mu}\psi_{A}$ is the covariant 
derivative w.r.t. $g_{\mu\nu}$). Then the stress tensor (the 
signature is $-+++$) 
\begin{equation}\label{n3}
T_{\mu\nu}(\psi_A)\equiv \frac{-2}{\sqrt{-g}}\frac{\delta}
{\delta g^{\mu\nu}}(L\sqrt{-g})
\end{equation}
vanishes identically and the spacetime evolves, in absence of 
other forms of matter, as an empty one, $G_{\mu\nu}=0$. We shall 
consider in the next section a few specific forms of $L_0$ for 
scalar and vector fields and find in which cases the corresponding 
Lagrange equations are trivial, making the fields unphysical. 
However it is clear that the conditions (\ref{n1}) and (2) are 
merely a (restrictive) sufficient condition for having 
$T_{\mu\nu}\equiv 0$ and not a necessary one. In fact, there are 
altogether 50 conditions (2) 
whereas there are only 10 identities $T_{\mu\nu}(\psi_A)
\equiv 0$ actually imposed on a possible Lagrangian. In what 
follows, we shall assume in Propositions 1 to 4 that the field 
Lagrangian $L(\psi_A, \psi_{A;\mu},g_{\mu\nu})$ may depend on 
the metric both explicitly and implicitly (via the covariant 
derivatives) and only in Examples 1 to 5 we shall consider 
various Lagrangians satisfying eqs. (1) and (2). \\

To avoid any confusion we emphasize that we work in the framework 
of classical field theory and thus we do not take into account 
the classical topological field theories \cite {FG} such as BF 
theory \cite{BBB}. They are metric independent and in this sense 
they might seem relevant to the present work, but their actions 
are typically given by surface integrals and they describe global 
observables related to the topological invariants of the manifolds 
on which they are defined. These theories have no local degrees 
of freedom, so there are no propagating field excitations 
(particles). \\

One may expect that if the system under consideration consists 
of a large number of fields, then vanishing of the full stress 
tensor will turn out to be a condition too weak to make trivial 
the large number of (coupled) Lagrange equations for the fields. 
We shall show on concrete examples that this is the case. Thus, 
we should theoretically allow for some specific systems of fields 
which carry no energy nor momentum, which 
nonetheless obey deterministic equations of motion (causal or 
not). The purpose of the present work is to show how many fields 
and of what type are necessary to this aim and how peculiar their 
Lagrangians must be. \\

The main thrust of the paper are Propositions 1 and 2 in sect. 2 
and Propositions 3 and 4 of sect. 3. Propositions 1 and 2 state 
that if a classical system consists either of a number of scalar 
fields or a single vector field and the stress tensor for the 
system is zero on and off shell, then Lagrange equations hold 
identically. This system is thus regarded as unphysical. 
However, if the number of scalar fields exceeds the dimensionality 
of the spacetime or there are two (or more) vector fields, the 
theorems break down. It is also relevant whether the dimensionality 
is even or odd. We show on specific examples that for sufficiently 
large number of fields (or their components) there exist 
Lagrangians satisfying conditions (\ref{n1}) and (\ref{n2}) and 
possessing the Lagrange equations of motion. Their dynamics is 
however quite bizarre: for systems of interacting fields there are 
no free-field solutions, for some cases the systems of propagation 
equations are either degenerate (of first order) or indeterministic 
(less equations than degrees of freedom). One concludes that 
classical fields without energy, though not excluded by principles 
of Lagrange field theory, require very peculiar and easily 
recognizable Lagrangians and these are unlikely to appear in 
modelling the physical reality. \\

Propositions 3 and 4 of sect. 3 solve, for interacting scalar 
fields or a single vector field, the problem of whether the stress 
tensor may vanish for all solutions while it is nonzero for some 
field configurations off shell. It turns out that there are always 
some solutions for which the stress tensor cannot vanish. In this 
sense nontrivial equations of motion imply in most cases a back 
reaction of the system on the spacetime. Unfortunately, for a 
vector field the proof works only in a small neighbourhood of flat 
spacetime and can be generalized solely to spacetimes admitting a 
covariantly constant Killing vector. \\
In section 4 we make some critical comments on the canonical 
energy--momentum tensor. Proofs of Propositions 1 and 2 employ the 
second Noether theorem and Proposition 4 requires the 
Belinfante--Rosenfeld identity. A detailed derivation of both the 
identities is provided in the Appendix. 

\section{Scalar and vector fields having no energy}
According to Introduction one should not expect to eliminate as 
unphysical the systems carrying no energy if they involve many 
interacting scalar and vector fields or fields with high spins. 
Instead one should separately investigate systems involving rather 
a small number of scalar fields or a single vector field. These 
fields are defined on a curved spacetime and are viewed either as 
test fields or as a part of a larger matter source of gravity. 
No specific gravitational field equations are assumed. The two 
systems are dealt with in the following two propositions. Let 
$d \geq 3$ be the dimensionality of the spacetime. \\

\textbf{Proposition 1} \\
Let a material system consist of $n\leq d$ interacting scalar 
fields and let the divergence of the stress tensor vanish  
identically for all values of the fields, $T^{\mu\nu}{}_{;\nu} 
\equiv 0$. Then their Lagrange equations of motion also hold 
identically for all values of the fields. \\

\textbf{Proposition 2} \\
Let the dimension $d$ be even and let a material system consist  
of a single vector field $A_{\mu}$ with a Lagrangian which can be 
expanded in a Taylor series in $A_{[\mu;\nu]}$ in the function space 
of all antisymmetric tensor fields $F_{\mu\nu}=-2A_{[\mu;\nu]}$ 
defined on an open domain in the spacetime. (The series is 
centered at $F_{\mu\nu}=0$ and thus the Lagrangian and all its 
derivatives are regular at this point.)
If the divergence of the stress tensor vanishes 
identically for all values of the vector field, $T^{\mu\nu}{}_{;\nu} 
\equiv 0$, then the equations of motion hold identically. 
Moreover, in 
$d=4$ the stress tensor itself is zero, $T^{\mu\nu}(A)\equiv 0$. \\

Idea of the proof. \\
The proof of both Propositions is based on the Noether identity, 
valid for any matter field, which arises from the coordinate 
invariance of the matter action integral. The identity is derived 
in the Appendix. It involves the divergence $T^{\mu\nu}{}_{;\nu}$ 
rather than the stress tensor itself and this is why the assumption 
of Propositions is apparently weaker than $T_{\mu\nu}\equiv 0$. 
At least in the case of a vector field and $d=4$ the two 
assumptions are equivalent. \\

\textbf{I. Proof of Proposition 1}\\
Consider a system of $n$ interacting scalar fields $\phi_a$, 
$a=1,\ldots, n$. 
For scalars the coefficients $Z_{A}{}^{\beta}{}_{\alpha}$ 
introduced in Appendix, eq. (A.15), are zero and the Noether  
identity (A.17) reduces to 
\begin{equation}\label{n4}
E^a\phi_{a;\mu}= T_{\mu\nu}{}^{;\nu},
\end{equation} 
where $T_{\mu\nu}$ depends on all the scalars and their first and 
second (in the case of a nonminimal coupling) order derivatives 
and the same holds for the $n$ scalar quantities 
$E^a$. (If $T_{\mu\nu}$ does involve $\phi_{a;\mu\nu}$, then the 
terms containing third order derivatives, arising on the r.h.s. of 
(4), do cancel each other.) 
For $n\leq d$ and 
arbitrary fields $\phi_a$, the vectors $\phi_{a;\mu}$ are linearly 
independent. On the other hand the condition $T_{\mu\nu}{}^{;\nu}
\equiv 0$ gives $E^a\phi_{a;\mu}\equiv 0$ suggesting 
that the gradients are actually linearly dependent. The consistency 
is restored only if 
$E^a\equiv 0$, i.e.~the field equations hold trivially. \\

The theorem breaks down for $n>d$ as is seen from the following 
example \footnote{Examples 1 and 4  were suggested to the author by Andrzej 
Staruszkiewicz.}. (Our experience with some readers shows that it should 
be explicitly stated that the determinant of any mixed tensor 
$Y^{\mu}{}_{\nu}$ is an absolute scalar, hence 
$\left|\det (Y_{\mu\nu})\right|^{1/2}$ is a scalar density, i.e. 
transforms as $\sqrt{-g}$. Accordingly, the Lagrangians in the Examples 
1, 4 and 5 are absolute scalars. On the other hand the Lagrangians in 
the Examples 2 and 3 are chosen as pseudoscalars (i.e. they transform 
as scalars multiplied by $J/|J|$, where $J$ is the Jacobian of a 
coordinate transformation) merely for 
computational simplicity and can be made scalars by taking the absolute 
value; this change will not affect the conclusions which follow from 
them.) \\

\textit{Example 1.} \\
For a system of $n$ scalar fields one defines 
$P_{\mu\nu}\equiv \sum_{a=1}^n \phi_{a,\mu}\phi_{a,\nu}$. For 
arbitrary scalars and $n\geq d$ its determinant 
$\det (P_{\mu\nu})\ne 0$ and may be used to make up a Lagrangian, 
specifically,
\begin{equation}\label{n5}
L_{\phi}=\frac{1}{\sqrt{-g}} 
\left|\det (P_{\mu\nu})\right|^{1/2}.
\end{equation} 
Then the conditions (\ref{n1})--(\ref{n2}) hold and 
$T_{\mu\nu}\equiv 0$. Let $d=4$ for simplicity. \\
i) For $n=4$ the Lagrangian takes on a simpler form, 
 \begin{displaymath}
L_{\phi}=\frac{1}{\sqrt{-g}} 
\left|\det (\phi_{a,\mu})\right|
\end{displaymath} 
and furthermore it is a full divergence,
\begin{equation}\label{n6}
L_{\phi}=\left|\nabla_{\alpha}(\varepsilon^{\alpha
\beta\mu\nu}
\phi_1 \phi_{2,\beta}\phi_{3,\mu}\phi_{4,\nu})\right|,
\end{equation}
where $\varepsilon ^{\alpha\beta\mu\nu}$ is the antisymmetric 
Levi-Civita pseudotensor (i.e. it transforms as a tensor times 
$J/|J|$) with $\varepsilon ^{0123}=1/\sqrt{-g}$. 
Clearly then $E^a(\phi_b)\equiv 0$. \\
ii) Yet in the case of $n=5$ scalars the Lagrangian (5) 
cannot be simplified to an analogous form and furthermore it is 
not a divergence. The five Euler operators do not vanish, 
\begin{eqnarray}\label{n7}
E^a(\phi_b)& \equiv & \frac{\partial L_{\phi}}
{\partial \phi_a}-\frac{1}{\sqg}\partial_{\mu}
\left(\sqg\frac{\partial L_{\phi}}{\partial \phi_{a,\mu}}
\right) 
\nonumber \\
& =& L_{\phi}Q^{\alpha\beta}(Q^{\mu\nu}\phi_{a,\nu}
\sum_{b=1}^{5}
\phi_{b,\mu}\phi_{b,\alpha\beta}-\phi_{a,\alpha\beta}),
\end{eqnarray}
where $Q^{\alpha\beta}$ is the symmetric inverse of $P_{\mu\nu}$, 
$Q^{\mu\alpha}P_{\alpha\nu}\equiv \delta^{\mu}_{\nu}$. To show 
that $E^a$ are not identically zero one finds counterexamples. In 
Minkowski spacetime one puts four of $\phi_a$ equal to the 
Cartesian coordinates and the fifth scalar equal to some nonlinear 
functions of time, e.g.~$t^2$ or $e^t$. Then two of the operators 
are different from zero. One concludes that the system of 5 scalar 
fields with the Lagrangian (\ref{n5}) is a form of 
non-self-gravitating and carrying no energy matter, subject to 
nonlinear propagation equations. The relationship 
$E^a\phi_{a,\mu}\equiv 0$ shows that there are at most four 
independent equations for the five scalars, thus the equations of 
motion and the initial data do not uniquely determine the 
evolution of the system. \\
The system might be detected, at least in principle, by its 
influence on the electromagnetic field $A_{\mu}$. Let 
$j^{\mu}\equiv q\sum_{a=1}^5 Q^{\mu\nu}\phi_{a,\nu}$, where $q$ 
is a coupling constant, be a current associated with the scalars. 
The current is coupled to the electromagnetic potential via the 
standard interaction term, $L_{\textrm{int}}=j^{\mu}A_{\mu}$, 
and the stress tensor corresponding to the Lagrangian 
$L_{\phi}+L_{\textrm{int}}$ remains identically zero. Yet the 
electromagnetic field is affected since the current enters the 
Maxwell equations. \\

\textbf{II. Proof of Proposition 2}\\
Here the main idea of the proof consists in expanding the Noether 
identity for the vector field, in the case where 
$T^{\mu\nu}{}_{;\nu} \equiv 0$, in a series of identities (not 
equations) ultimately resulting in the Euler operator which 
vanishes identically. \\

For a single vector field the Noether identity (A.17) takes the 
form (A.18),
 \begin{equation}\label{n8}
E^{\mu}F_{\alpha\mu}\equiv T_{\alpha\nu}{}^{;\nu}+
A_{\alpha}E^{\mu}{}_{;\mu}
\end{equation} 
with $F_{\alpha\beta}\equiv A_{\beta;\alpha}-A_{\alpha;\beta}$ 
being a "field strength" and the Euler operator 
\begin{equation}\label{n9}
E^{\mu}[L(A)]\equiv \frac{\partial L}{\partial A_{\mu}}-
\nabla_{\alpha}\left(\frac{\partial L}{\partial A_{\mu;\alpha}}
\right).
\end{equation} 
We set $T_{\alpha\nu}{}^{;\nu}\equiv 0$ and decompose the 
identity (8) into a sum of terms containing covariant 
derivatives of $A_{\mu}$ of definite order. To this aim we first 
introduce tensors 
\begin{equation}\label{n10}
B^{\mu\gamma\alpha\beta}\equiv \frac{\partial^2 L}{\partial 
A_{\mu;\gamma}\partial A_{\alpha;\beta}}=
B^{\alpha\beta\mu\gamma},
\end{equation} 
\begin{equation}\label{n11}
d^{\alpha\beta}\equiv \frac{\partial L}{\partial A_{\alpha;
\beta}}\qquad \textrm{and}\qquad l^{\mu}\equiv 
\frac{\partial L}{\partial A_{\mu}},
\end{equation} 
these are functions of $A_{\mu}$ and $A_{\mu;\nu}$ (and the 
metic). Then one finds 
\begin{equation}\label{n12}
E^{\mu}=-B^{\mu\gamma\alpha\beta} 
A_{\alpha;\beta\gamma}-\frac{\partial d^{\mu\alpha}}
{\partial A_{\nu}}A_{\nu;\alpha}+l^{\mu}.
\end{equation} 
We insert this expression for $E^{\mu}$ into (8) and 
decrease the order of derivatives of $A_{\mu}$ with the aid of 
Ricci identity. After some manipulations the identity 
$E^{\mu}F_{\alpha\mu} -
A_{\alpha}E^{\mu}{}_{;\mu}\equiv 0$ takes on the following 
involved form,
\begin{eqnarray}\label{n13}
0 & \equiv & A_{\alpha}B^{\mu\nu\beta\gamma} 
A_{\beta;\gamma(\mu\nu)}+ A_{\alpha}\frac{\partial 
B^{\mu\gamma\nu\beta}}{\partial A_{\lambda;\sigma}}
A_{\lambda;(\mu\sigma)}A_{\nu;(\beta\gamma)}+
\nonumber \\
& & A_{\lambda;(\mu\sigma)}\left\{-F_{\alpha\nu}B^{\nu\sigma
\lambda\mu}+ A_{\alpha}\left[A^{\tau}R_{\tau\nu\beta\gamma}
\frac{\partial B^{\lambda\sigma(\mu\gamma)}}{\partial 
A_{\nu;\beta}} +2\frac{\partial B^{(\mu\nu)\lambda\sigma}}
{\partial A_{\beta}} +\frac{\partial d^{(\mu\sigma)}}
{\partial A_{\lambda}} -\frac{\partial d^{\lambda\sigma}}
{\partial A_{\mu}}\right]\right\}
\nonumber \\
& & +a_{\alpha}(A_{\mu}, A_{\mu;\nu}, R_{\mu\nu\lambda\sigma}),
\end{eqnarray}
where $a_{\alpha}$ is a complicated term containing no higher 
derivatives than of first order. Being an identity for arbitrary 
$A_{\mu}$, the terms with derivatives of different order cannot 
cancel each other, instead they should vanish separately. This 
implies that the identity splits into a cascade of four sets of 
independent identities. We first consider the term with third 
derivatives. It vanishes if their coefficients are zero. Since 
$A_{\alpha}\ne 0$ one gets $B^{(\mu\nu)\beta\gamma}\equiv 0$, 
then the definition (10) yields 
\begin{eqnarray}\label{n14}
B^{\alpha\beta\mu\nu}=B^{\mu\nu\alpha\beta}=-B^{\beta\alpha\mu\nu}
=-B^{\alpha\beta\nu\mu}.
\end{eqnarray}
These in turn imply a specific relationship between 
$B^{\alpha\beta\mu\nu}$ and $d^{\alpha\beta}$. It turns out 
useful to decompose $A_{\alpha;\beta}$ and $d^{\alpha\beta}$ into 
symmetric and antisymmetric parts, $A_{\alpha;\beta}=
A_{(\alpha;\beta)}+\frac{1}{2}F_{\beta\alpha}$ and 
\begin{eqnarray}\label{n15}
d^{(\alpha\beta)}=\frac{\partial L}{\partial A_{(\alpha;
\beta)}}\equiv S^{\alpha\beta}, \qquad d^{[\alpha\beta]}=
2\frac{\partial L}{\partial F_{\beta\alpha}}\equiv 
N^{\alpha\beta}.
\end{eqnarray}
Since
\begin{displaymath}
B^{\mu\nu\alpha\beta}=\frac{\partial d^{\alpha\beta}}
{\partial A_{\mu;\nu}}=\frac{\partial d^{\mu\nu}}
{\partial A_{\alpha;\beta}}
\end{displaymath}
one finds 
\begin{eqnarray}\label{n16}
B^{\alpha\beta\mu\nu}=\frac{\partial S^{\alpha\beta}}
{\partial A_{(\mu;\nu)}} +\frac{\partial N^{\alpha\beta}}
{\partial A_{(\mu;\nu)}} -2\frac{\partial S^{\alpha\beta}}
{\partial F_{\mu\nu}} -2\frac{\partial N^{\alpha\beta}}
{\partial F_{\mu\nu}}
\end{eqnarray}
and a similar expression with the pairs $\alpha\beta$ and 
$\mu\nu$ interchanged. By antisymmetrizing in these pairs one 
gets
\begin{eqnarray}\label{n17}
B^{[\alpha\beta][\mu\nu]}=-2\frac{\partial N^{\alpha\beta}}
{\partial F_{\mu\nu}}=-2\frac{\partial N^{\mu\nu}}
{\partial F_{\alpha\beta}}.
\end{eqnarray}
On the other hand the symmetries (14) mean that 
$B^{\alpha\beta\mu\nu}=B^{[\alpha\beta][\mu\nu]}$ and by 
equating eq. (16) to (17) one arrives at the 
following restrictions imposed on $d^{\alpha\beta}$,
\begin{eqnarray}\label{n18}
\frac{\partial S^{\alpha\beta}}{\partial A_{(\mu;\nu)}}=
\frac{\partial S^{\alpha\beta}}{\partial F_{\mu\nu}}= 
\frac{\partial N^{\alpha\beta}}{\partial A_{(\mu;\nu)}}=0.
\end{eqnarray}
From these one infers that the Lagrangian may depend on 
$A_{(\alpha;\beta)}$ only via linear terms, 
\begin{eqnarray}\label{n19}
L=f_1(s)A^{\alpha}A^{\beta}A_{(\alpha;\beta)}+
f_2(s)g^{\alpha\beta}A_{(\alpha;\beta)}+
L_2(A_{\mu}, F_{\mu\nu}),
\end{eqnarray}
where $f_1$ and $f_2$ are arbitrary smooth functions of the 
vector length, $s\equiv A^{\mu}A_{\mu}$. The first two terms in 
(19) can be expressed as $\nabla_{\alpha}(f(s)A^{\alpha})+
h(s)A^{\alpha}_{;\alpha}$, with $2\ud f/\ud s\equiv f_1$ and 
$h\equiv f_2-f$. Then discarding the divergence term from the 
Lagrangian one finds
\begin{eqnarray}\label{n20}
d^{\alpha\beta}=hg^{\alpha\beta}+N^{\alpha\beta} \qquad
\textrm{and} \qquad 
B^{\alpha\beta\mu\nu}=-2\frac{\partial N^{\alpha\beta}}
{\partial F_{\mu\nu}}=-2\frac{\partial N^{\mu\nu}}
{\partial F_{\alpha\beta}}.
\end{eqnarray}
We next study the identities involving $A_{\nu;(\beta\gamma)}$ 
quadratically. From (13) one sees that vanishing of the 
coefficients of these terms requires 
\begin{eqnarray}\label{n21}
\frac{\partial B^{\mu(\beta\gamma)\nu}}{\partial 
F_{\sigma\lambda}} +\frac{\partial B^{\sigma(\beta\gamma)\nu}}
{\partial F_{\mu\lambda}} \equiv 0.
\end{eqnarray}
By employing eqs. (20), however, one cannot simplify these 
identities (after inserting (20) into (21) one recovers the 
identities with the pairs of indices $\beta\gamma$ and 
$\mu\sigma$ interchanged), therefore one passes to studying the 
identities involving $A_{\lambda;(\mu\sigma)}$ linearly. These 
imply vanishing of the curly bracket in eq. (13). Making use of 
(20) this reads 
\begin{eqnarray}\label{n22}
F_{\alpha\nu}B^{\nu(\mu\sigma)\lambda}+ A_{\alpha}
C^{\mu\sigma\lambda}\equiv 0
\end{eqnarray}
with
\begin{eqnarray}\label{n23}
C^{\mu\sigma\lambda}\equiv 2h'(g^{\mu\sigma}A^{\lambda}-
A^{(\mu}g^{\sigma)\lambda})- \frac{1}{2}
\left(\frac{\partial N^{\lambda\sigma}}{\partial A_{\mu}}+
\frac{\partial N^{\lambda\mu}}{\partial A_{\sigma}}\right)=
C^{(\mu\sigma)\lambda}
\end{eqnarray}
and $h'=\ud h/\ud s$. All the time one investigates a generic 
field $A_{\mu}$ for which $\det(F_{\mu\nu})\ne 0$. (It is here 
that the even number of spacetime dimensions becomes relevant; 
for $d$ odd, $\det(F_{\mu\nu})= 0$.) Then there exists the 
inverse  matrix $f^{\alpha\beta}=-f^{\beta\alpha}$ given by 
$f^{\alpha\mu}F_{\mu\beta}=F_{\beta\mu}f^{\mu\alpha}=\delta
^{\alpha}_{\beta}$. Multiplying (22) by $f^{\alpha\tau}$ one 
gets 
\begin{eqnarray}\label{n24}
B^{\alpha(\mu\nu)\beta}=-f^{\alpha\sigma}A_{\sigma}
C^{\mu\nu\beta}.
\end{eqnarray}
One immediately sees that this expression implies $A_{\alpha}
B^{\alpha(\mu\nu)\beta}\equiv 0$. On the other hand one infers 
from (20) that $B^{\alpha(\mu\nu)\beta}\equiv 
B^{\beta(\mu\nu)\alpha}$. Applying this symmetry to the former 
identity one finds 
\begin{eqnarray}\label{n25}
B^{\alpha(\mu\nu)\beta}A_{\beta}=-f^{\alpha\sigma}A_{\sigma}
C^{\mu\nu\beta}A_{\beta}\equiv 0.
\end{eqnarray}
This in turn implies that $C^{\mu\nu\beta}A_{\beta}\equiv 0$ 
since for a generic vector field $f^{\alpha\sigma}A_{\sigma}
\ne 0$. By inspection of the expression (23) one concludes 
that the term proportional to $h'$ in the latter identity 
contains no derivatives, whereas the other term in this identity 
is a sum of terms each of which does contain $F_{\mu\nu}$. In 
fact, terms of the form $\partial N^{\mu\nu}/\partial A_
{\alpha}$ cannot involve an additive term free of the 
derivatives, since the latter would only arise from a linear 
term in the Lagrangian, $k^{\mu\nu}(A)F_{\mu\nu}$. However, 
for $d$ even, an antisymmetric $k^{\mu\nu}$ cannot be made up 
alone of $A_{\mu}$, 
$g_{\mu\nu}$ and the Levi--Civita tensor. In conclusion, the 
identity $C^{\mu\nu\beta}A_{\beta}\equiv 0$ splits into two 
sets, 
\begin{eqnarray}\label{n26}
h'(g^{\mu\nu}s-A^{(\mu}A^{\nu)})\equiv 0
\end{eqnarray}
\begin{eqnarray}\label{n27}
\textrm{and} \qquad A_{\beta}
\left(\frac{\partial N^{\beta\nu}}{\partial A_{\mu}}+
\frac{\partial N^{\beta\mu}}{\partial A_{\nu}}\right)
\equiv 0.
\end{eqnarray}
Eq. (26) may be satisfied only if $h=h_0=const$; then 
$h_0 A^{\alpha}_{;\alpha}$ is discarded as being a full 
divergence and finally $L=L(A_{\mu}, F_{\mu\nu}, g_{\mu
\nu})$. \\

We now return to investigating the expression (24). It 
is very peculiar. First, the r.h.s. of the expression does not 
have the above mentioned symmetry $B^{\alpha(\mu\nu)\beta}\equiv 
B^{\beta(\mu\nu)\alpha}$. Second, according to the assumption 
of the theorem, $B^{\alpha\mu\nu\beta}$ is analytic (is a 
Taylor series) in 
$F_{\mu\nu}$ about $F_{\mu\nu}=0$ and the same holds for 
$C^{\mu\nu\beta}$. Yet $B^{\alpha(\mu\nu)\beta}$ is proportional 
to $f^{\alpha\sigma}A_{\sigma}$ and the expression should be 
valid also about $F_{\mu\nu}=0$ where the inverse 
$f^{\alpha\sigma}$ does not exist. (Actually the expression 
breaks down whenever $\det(F_{\mu\nu})= 0$.) The dependence on 
$f^{\alpha\sigma}$ cannot be eliminated since there is no 
contraction of this tensor with any of $F_{\mu\nu}$ appearing 
in $C^{\mu\nu\beta}$. One then infers that identity (24) may 
hold only if $B^{\alpha(\mu\nu)\beta}\equiv 0$. \\
To avoid any confusion it should be emphasized that the above 
reasoning does not apply to the Noether identity (8). In fact, 
with the aid of $f^{\alpha\mu}$ it can be reexpressed as 
\begin{displaymath}
E^{\alpha}=f^{\alpha\mu}(A_{\mu}E^{\nu}_{;\nu}+T_{\mu\nu}{}
^{;\nu})
\end{displaymath}
and apparently $E^{\alpha}$ does explicitly depend on 
$f^{\alpha\mu}$, contrary to the assumption of the theorem. 
However, here $f^{\alpha\mu}$ is contracted with $T_{\mu\nu}{}
^{;\nu}$ and the original form of the identity ensures that the 
$f^{\alpha\mu}$--dependence is trivially cancelled. This fact 
stresses the role played by $T_{\mu\nu}$. \\

The identity $B^{\alpha(\mu\nu)\beta}\equiv 0$ makes identities 
(21) trivial and has two further consequences. First, together 
with (14) and (20) it implies that $B^{\alpha\mu\nu\beta}$ is 
totally antisymmetric,
\begin{eqnarray}\label{n28}
B^{\alpha\mu\nu\beta}=B^{[\alpha\mu\nu\beta]}.
\end{eqnarray}
In $d=4$ one then infers that $B^{\alpha\mu\nu\beta}=
p(A_{\lambda}, F_{\lambda\sigma})\varepsilon^{\alpha\mu\nu
\beta}$ with some definite pseudoscalar function $p$. Second, it 
requires $C^{\mu\nu\beta}\equiv 0$ or from (23),
\begin{eqnarray}\label{n29}
\frac{\partial N^{\alpha\mu}}{\partial A_{\nu}}+
\frac{\partial N^{\alpha\nu}}{\partial A_{\mu}}
\equiv 0.
\end{eqnarray}
The latter shows that the following tensor is totally 
antisymmetric, 
\begin{eqnarray}\label{n30}
n^{\alpha\mu\nu}\equiv \frac{\partial N^{\alpha\mu}}
{\partial A_{\nu}}=n^{[\alpha\mu\nu]}.
\end{eqnarray}
The formula (12) for the Euler operator is reduced, upon 
applying eqs. (28) and (30), to 
\begin{eqnarray}\label{n31}
E^{\mu}=-\frac{1}{2}n^{\mu\alpha\beta}F_{\alpha\beta}+
l^{\mu},
\end{eqnarray}
now it contains no second order derivatives. \\

Finally we study the last system of identities arising from 
(13), those involving at most the first derivatives of $A_
{\mu}$, i.e. $a_{\alpha} \equiv 0$. After some manipulations 
with the use of the Riemann tensor symmetries and eqs. (28) and 
(30) one arrives at the following expression:
\begin{eqnarray}\label{n32}
a_{\alpha}=F_{\alpha\mu}E^{\mu} -
A_{\alpha}\frac{\partial E^{\mu}}{\partial A_{\nu}}
A_{\nu;\mu}\equiv 0
\end{eqnarray}
with $E^{\mu}$ given by (31). As in the case of 
$B^{\alpha(\mu\nu)\beta}$ one may use the analyticity property 
to prove that $E^{\mu}\equiv 0$. It is interesting, however, to 
see that this result can also be attained in an independent way. 
Using the formula (31) one easily finds that the scalar in the 
last term of (32) is equal to 
\begin{eqnarray}\label{n33}
\frac{\partial E^{\mu}}{\partial A_{\nu}}A_{\nu;\mu}=
\frac{\partial^2 L}{\partial A_{\mu}\partial A_{\nu}} 
A_{\mu;\nu}.
\end{eqnarray}
One sees that the r.h.s. of eq. (33) depends linearly on 
$A_{(\mu;\nu)}$ and upon inserting the scalar back into eq. (32) 
the term $F_{\alpha\mu}E^{\mu}$ acquires the same dependence. On 
the other hand it has already been proved that $L=L(A_{\mu}, 
F_{\mu\nu}, g_{\mu\nu})$ and the symmetrized derivative cannot 
arise in the process of differentiation of the Lagrangian. The 
contradiction is removed by requiring that
\begin{eqnarray}\label{n34}
\frac{\partial^2 L}{\partial A_{\mu}\partial A_{\nu}}\equiv 0
\end{eqnarray}
or $L=L^{\mu}(F_{\alpha\beta})A_{\mu} +L_0(F_{\alpha\beta})$. 
However for even number of dimensions it is impossible to make 
up a vector out of $F_{\alpha\beta}$, $g_{\mu\nu}$ and the 
Levi--Civita tensor. In consequence $L^{\mu}=0$ and the 
Lagrangian is some function of $F_{\alpha\beta}$ and 
$g_{\mu\nu}$ alone. \\
As a result of vanishing of the scalar (33) one gets from (32) 
that $F_{\alpha\mu}E^{\mu}\equiv 0$ and finally one arrives at 
the conclusion that for a generic vector field the Euler 
operator associated with the Lagrangian generating the stress 
tensor with $T_{\mu\nu}{}^{;\nu}\equiv 0$ vanishes identically, 
$E^{\mu}[L(A)]\equiv 0$. \\
This outcome is in agreement with eq. (31) since the Lagrangian 
does not depend on $A_{\mu}$ and then $n^{\alpha\mu\nu}\equiv 0 
\equiv l^{\mu}$. \\

The last step of the proof consists in finding a relationship 
between the identity $T_{\mu\nu}{}^{;\nu}\equiv 0$ and the 
stress tensor itself in the case $d=4$. From eq. (28) one 
infers that 
\begin{eqnarray}\label{n35}
B^{\alpha\mu\nu\beta}=4\frac{\partial^2 L}
{\partial F_{\alpha\mu}\partial F_{\nu\beta}}=
p(F_{\lambda\sigma})\varepsilon^{\alpha\mu\nu\beta}
\end{eqnarray}
with unknown pseudoscalar $p$. The Lagrangian, which depends on 
$A_{\alpha}$ only via $F_{\alpha\beta}$, is a function of the 
two invariants of the field strength, 
$V=F_{\alpha\beta}F^{\alpha\beta}$ and $W=P^2$ where 
$P=\varepsilon^{\alpha\beta\mu\nu}F_{\alpha\beta}F_{\mu\nu}$ 
is a pseudoscalar, i.e. $L=L(W,V)$. Then 
\begin{eqnarray}\label{n36} 
B^{\alpha\mu\nu\beta} & = & 32PL_{WV}(\varepsilon^
{\alpha\mu\lambda\sigma}F^{\nu\beta}+ F^{\alpha\mu}
\varepsilon^{\nu\beta\lambda\sigma})F_{\lambda\sigma} +
32(L_W+2WL_{WW})\varepsilon^{\alpha\mu\lambda\sigma}F_{\lambda\sigma}
\varepsilon^{\nu\beta\tau\rho}F_{\tau\rho} 
\nonumber\\
& & +16L_{VV}F^{\alpha\mu}F^{\nu\beta} +
4L_V(g^{\alpha\nu}g^{\mu\beta}-g^{\alpha\beta}
g^{\mu\nu}) + 16PL_W\varepsilon^{\alpha\mu\nu\beta}
\nonumber\\
& \equiv & J^{\alpha\mu\nu\beta}+ 
16PL_W\varepsilon^{\alpha\mu\nu\beta}=
p(W, V)\varepsilon^{\alpha\mu\nu\beta},
\end{eqnarray}
where $L_V=\partial L/\partial V$ etc. The first four terms 
have lower symmetry than that required: only $\alpha\mu\nu
\beta=[\alpha\mu][\nu\beta]=\nu\beta\alpha\mu$, thus their sum 
$J^{\alpha\mu\nu\beta}$  must vanish identically and this is 
possible only if each of them 
is separately zero. To see this one assumes that the sum is zero 
for a fixed field $F_{0\mu\nu}$. Let $F_{\mu\nu}=
F_{0\mu\nu}+\delta F_{\mu\nu}$ where $\delta F_{\mu\nu}$ is 
arbitrary infinitesimal. The variation 
$\delta J^{\alpha\mu\nu\beta}$ also must vanish,
\begin{eqnarray}\label{n37} 
\delta J^{\alpha\mu\nu\beta} & = & 32\delta(PL_{WV})
(\varepsilon^{\alpha\mu\lambda\sigma}F_{0}^{\nu\beta}+ 
F_{0}^{\alpha\mu}\varepsilon^{\nu\beta\lambda\sigma})
F_{0\lambda\sigma} +
\nonumber\\
& & 
32PL_{WV}[(\varepsilon^{\alpha\mu\lambda\sigma}\delta F^{\nu\beta}
+ \delta F^{\alpha\mu}
\varepsilon^{\nu\beta\lambda\sigma})F_{0\lambda\sigma}+
(\varepsilon^{\alpha\mu\lambda\sigma}F_{0}^{\nu\beta}
+ F_{0}^{\alpha\mu}
\varepsilon^{\nu\beta\lambda\sigma})\delta F_{\lambda\sigma}]
\nonumber\\
& & 
+32\delta(L_W+2WL_{WW})\varepsilon^{\alpha\mu\lambda\sigma}
F_{0\lambda\sigma}\varepsilon^{\nu\beta\tau\rho}F_{0\tau\rho} 
\nonumber\\
& & 
+32(L_W+2WL_{WW})\varepsilon^{\alpha\mu\lambda\sigma}
\varepsilon^{\nu\beta\tau\rho}(\delta F_{\lambda\sigma}
F_{0\tau\rho}+ F_{0\lambda\sigma}\delta F_{\tau\rho})
 +16\delta L_{VV}F_{0}^{\alpha\mu}F_{0}^{\nu\beta} 
\nonumber\\
& & 
+16 L_{VV}(\delta F^{\alpha\mu}F_{0}^{\nu\beta}+
F_{0}^{\alpha\mu}\delta F^{\nu\beta})+ 
4\delta L_V(g^{\alpha\nu}g^{\mu\beta}-g^{\alpha\beta}
g^{\mu\nu}) \equiv 0.
\end{eqnarray}
Here for any $\Phi(W,V)$ one has as usual $\delta \Phi=
\Phi_W\delta W+\Phi_V\delta V$ where $\delta W=2P_0\delta P
=4P_0\varepsilon^{\alpha\beta\mu\nu}F_{0\alpha\beta}
\delta F_{\mu\nu}$ and $\delta V= 2F_{0}^{\mu\nu}
\delta F_{\mu\nu}$, all the derivatives and $W$ and $V$ 
are taken at $F_{0\mu\nu}$. At fixed $F_{0\mu\nu}$ one 
varies $\delta F_{\mu\nu}$ in such a way that the variations 
$\delta W$ and $\delta V$ remain unaltered. Then the first, 
third, fifth and seventh (the last) terms in (37) remain 
constant whereas the other three are variable. For the 
identity holds for any $\delta F_{\mu\nu}$, the constant 
terms must vanish, i.e. $\delta(PL_{WV})=\delta(L_W+
2WL_{WW})=\delta L_{VV}=\delta L_V=0$ and these identities 
hold for any values of $\delta W$ and $\delta V$. The last 
identity yields then $L_{VW}=L_{VV}=0$ or $L(W,V)=L(W)+ aV$, 
$a$ is constant. Inserting this Lagrangian into eq. (36) 
one sees that the first and third terms are zero whereas the 
fourth term becomes independent of $F_{\mu\nu}$, thus it 
must vanish too. This implies $L_V=0$ or $a=0$. The 
vanishing sum $J^{\alpha\mu\nu\beta}=0$ consists now of the 
second term alone and it is zero provided $L_W+2WL_{WW}=0$. 
A general solution to this equation is (dropping an additive 
constant) $L=c\sqrt{W}=c|P|$ with constant $c$. Finally,
\begin{displaymath}
L=2c|\nabla_{\alpha}(\varepsilon^{\alpha\beta\mu\nu}A_{\beta}
F_{\mu\nu})|,
\end{displaymath}
a textbook result \cite{LL}.\\
This completes the proof of Proposition 2. \\

For dimensions $d=2n$, $n>2$, the tensor $F_{\alpha\beta}$ has 
more invariants. One may conjecture that also in higher 
dimensions eq. (28) implies that 
\begin{displaymath}
L=c |\varepsilon^{\alpha_{1}\beta_{1}\ldots\alpha_{n}\beta_{n}}
F_{\alpha_{1}\beta_{1}}\ldots F_{\alpha_{n}\beta_{n}}|
\end{displaymath}
being a full divergence and hence $T_{\mu\nu}(A)\equiv 0$. 
However it is harder to prove that this solution is unique. \\

In principle one might envisage a scalar (or $n<d$ scalars) or 
a single vector field in $d>4$ even with $T_{\mu\nu}\neq 0$ and 
$T^{\mu\nu}{}_{;\nu}\equiv 0$. Formally such a field might appear 
in  Einstein field equations. However, since  
the Lagrange equations of motion are trivial and the field has 
no determined propagation, it should be rejected on physical 
grounds. At first sight the stress tensor for a ground state 
solution of a quantum field, $<T_{\mu\nu}>=\rho_V g_{\mu\nu}$, 
where $\rho_V$ is a constant energy density for the classical 
nonzero value of the quantum field in this state, contradicts the 
above statement. It should be therefore emphasized that this 
expectation value of $T_{\mu\nu}$ comes from semiclassical 
considerations 
and cannot be derived within classical field theory (formally 
this expression is generated by the cosmological constant term 
in the full Lagrangian including gravity); the proposition does 
not apply to that case. \\ 

The Proposition 2 cannot be generalized neither to odd number 
of dimensions nor to more than one vector field. This is shown 
by the following counterexamples. \\

\textit{Example 2.} \\
Let $d=3$. If one chooses a pseudoscalar 
$L=\varepsilon^{\alpha\beta\gamma}F_{\alpha\beta}A_{\gamma}$, 
then $T_{\mu\nu}(A)\equiv 0$. This 
Lagrangian is not a divergence and $E^{\mu}=2\varepsilon^
{\mu\alpha\beta}F_{\alpha\beta}$, the operator involves no 
second order derivatives since $L$ is degenerate being linear 
in $F_{\alpha\beta}$. The field equations read then 
$F_{\alpha\beta}=0$ and admit only one solution, the vacuum. 
The model is trivial. \\

\textit{Example 3.} \\
In $d=5$ the analogous model is nontrivial. Let 
\begin{eqnarray}\label{n38}
L=W^{\mu}A_{\mu}=\varepsilon^{\mu\alpha\beta\gamma\delta}
F_{\alpha\beta}F_{\gamma\delta}A_{\mu},
\end{eqnarray}
where $W^{\mu}\equiv \varepsilon^{\mu\alpha\beta\gamma\delta}
F_{\alpha\beta}F_{\gamma\delta}$. The stress tensor is zero, 
$T_{\mu\nu}(A)\equiv 0$, whereas the Lagrangian is not a 
divergence and is gauge invariant under $A'_{\mu}=A_{\mu}+
\partial_{\mu}\chi$. $L$ is apparently nondegenerate as being 
quadratic in $F_{\alpha\beta}$, nevertheless the Euler operator 
is of first order, $E^{\mu}=3W^{\mu}$, and in this sense the 
propagation equations for the field are degenerate. One finds 
that $E^{\mu}_{;\mu}\equiv 0$ owing to the gauge invariance. 
Then the Noether identity (8) reduces to $F_{\alpha\mu}E^{\mu}
\equiv 0$ with $E^{\mu}$ nonvanishing in general ($\det
(F_{\mu\nu})\equiv 0$). The field equations $W^{\mu}=0$ are 
quadratic in $F_{\alpha\beta}$ and admit nonzero solutions. A 
particular solution is, e.g.
\begin{eqnarray}\label{n39}
A_1=a_1 x^0-a_2 x^2-a_3 x^3 -a_4 x^4, \qquad 
A_0=A_2=A_3=A_4=0,
\end{eqnarray}
yielding $F_{01}=a_1$, $F_{12}=a_2$, $F_{13}=a_3$ and $F_{14}=
a_4$, otherwise zero, with $a_1,\ldots,a_4$ constant. \\

\textit{Example 4.} \\
Next we consider two vector fields forming an open system, $d=
4$. A field $A_{\mu}$ interacts with a given external field 
$W^{\mu}$. First one defines an antisymmetric tensor 
$V_{\mu\nu}\equiv W^{\alpha}A_{[\alpha}\partial_{\mu}
A_{\nu]}=\frac{1}{2}W^{\alpha}A_{[\alpha}F_{\mu\nu]}$, where as 
usual, $F_{\mu\nu}=A_{\nu,\mu}-A_{\mu,\nu}$. The tensor is 
metric independent and thus may be used to constructing an 
interaction Lagrangian,  
\begin{eqnarray}\label{n40}
L(A,W) & \equiv & \frac{1}{\sqg}[-\det(V_{\mu\nu})]^{1/2}=
\frac{1}{\sqrt{8}}[2V_{\alpha\beta}V^{\beta\mu}V_{\mu\nu}
V^{\nu\alpha}-(V_{\mu\nu}V^{\mu\nu})^2]^{1/2}
\nonumber\\
& & =\frac{1}{8}|\varepsilon^{\alpha\beta\mu\nu}V_{\alpha\beta}
V_{\mu\nu}|.
\end{eqnarray}
$L$ does not involve derivatives of the external field. In this 
model there is no free Lagrangian for $A_{\mu}$ and obviously 
$L(A,0)\equiv 0$. The stress tensor generated by the interaction 
Lagrangian is $T_{\mu\nu}(A,W)\equiv 0$. On the other hand if one 
attempts to complete the system, i.e.~to make it closed, by 
addding a Lagrangian $L_W$ for $W^{\mu}$ (free or including an 
interaction term), then $\sqg L_W$ must depend on the metric. 
Thus a full stress tensor for a closed system of the two vector 
fields is different from zero. \\
   The open system described by $L$ as in (40) has no simple 
symmetries. The Euler operator $E^{\mu}$ for $A_{\mu}$ is 
nondegenerate and very complicated. The second order derivatives 
appear in it in a term of the form 
\begin{eqnarray}\label{n41}
\varepsilon^{\mu\nu\alpha\beta}A_{\nu}A_{\alpha;(\beta\gamma)}
W^{\gamma}W^{\sigma}A_{\sigma}.
\end{eqnarray}

\textit{Example 5.} \\
One can construct a model for a closed system of two interacting 
vector fields in $d=4$. One assigns field strengths to vector 
potentials $A_{\mu}$ and $B_{\mu}$:
\begin{eqnarray}\label{n42}
F_{\mu\nu}=A_{\nu,\mu}-A_{\mu,\nu} \qquad \textrm{and} 
\qquad H_{\mu\nu}=B_{\nu,\mu}-B_{\mu,\nu}.
\end{eqnarray}
The interaction Lagrangian is chosen as 
\begin{eqnarray}\label{n43}
L(A,B)\equiv \frac{1}{\sqg}[\det(F_{\mu\nu})\det(H_{\mu\nu})]
^{1/4}=\frac{1}{8}\vert L_A(F)L_B(H)\vert^{1/2},
\end{eqnarray}
where $L_A\equiv \varepsilon^{\alpha\beta\mu\nu}
F_{\alpha\beta}F_{\mu\nu}$ and $L_B\equiv \varepsilon^
{\alpha\beta\mu\nu}H_{\alpha\beta}H_{\mu\nu}$ are pseudoscalars. 
As in Example 4 
there are no free Lagrangians for the fields (by introducing free 
Lagrangians one will render the full Lagrangian density metric 
dependent, Proposition 2) and the interaction Lagrangian is 
gauge invariant under $A'_{\mu}=A_{\mu}+\partial_{\mu}\chi_1$ 
and $B'_{\mu}=B_{\mu}+\partial_{\mu}\chi_2$, $\chi_1$ and 
$\chi_2$ arbitrary. The interaction energy and momentum for the 
system are zero, $T_{\mu\nu}(A,B)\equiv 0$, whereas the Euler 
operator consists of two vector ones, 
\begin{eqnarray}\label{n44}
E^{\mu}_{A}\equiv \frac{\delta L}{\delta A_{\mu}}=
-\frac{L}{L_A}\varepsilon^{\mu\nu\alpha\beta}F_{\alpha\beta}
\left(\frac{L_{A;\nu}}{L_A}- \frac{L_{B;\nu}}{L_B}\right)
\end{eqnarray}
and analogously, by applying the symmetry of interchanging the 
fields in $L$, 
\begin{eqnarray}\label{n45}
E^{\mu}_{B}\equiv \frac{\delta L}{\delta B_{\mu}}=
-\frac{L}{L_B}\varepsilon^{\mu\nu\alpha\beta}H_{\alpha\beta}
\left(\frac{L_{B;\nu}}{L_B}- \frac{L_{A;\nu}}{L_A}\right). 
\end{eqnarray}
The operators contain second order derivatives (are 
nondegenerate). One can find a special solution to the field 
equations $E^{\mu}_{A}=0=E^{\mu}_{B}$ in Minkowski spacetime in 
the following way. One assumes that the "electric" and "magnetic" 
fields corresponding to $F_{\mu\nu}$ are parallel and of equal 
length and direction, then $F^{\mu\nu}F_{\mu\nu}=0$ and $L_A=
-8a^2$, $a>0$ with $F_{01}=F_{32}=a$, otherwise zero; and 
analogously, $H^{\mu\nu}H_{\mu\nu}=0$ and $L_B=-8b^2$, $b>0$ and 
$H_{01}=H_{32}=b$. In this case the two operators are 
proportional, 
\begin{eqnarray}\label{n46}
E^{\mu}_{A}=\frac{b}{2}(-\partial_1, \partial_0, \partial_3, 
-\partial_2)\ln \frac{a}{b}=-\frac{a}{b}E^{\mu}_{B}.
\end{eqnarray} 
A particular solution is then $A_{\mu}=-a(x, 0, 0, y)$ and 
$B_{\mu}=-b(x, 0, 0, y)$ with $a$ and $b$ constant. Eq. (46) 
shows that $a$ and $b$ cannot vanish (no free fields). 

\section{The stress tensor vanishing for solutions}
The Propositions 1 and 2 raise an obvious problem: is it 
possible that $T_{\mu\nu}$, while being different from zero for 
some values ("off shell") of the field under consideration, 
$\psi_A$, vanishes for all solutions of 
$E^A=0$ and not only for the trivial solution $\psi_A=0$? Such 
a possibility would be more disturbing than the case 
$T_{\mu\nu}\equiv 0$ for all values of the field, for field 
configurations which do not obey some equations of motion do not 
exist in nature. Such a matter, subject to deterministic causal 
propagation equations, would be a truly nongravitating one. In 
the framework of classical field theory one can almost exclude 
such fields under reasonable assumptions. We again consider only 
scalar fields and a single vector field. \\

\textbf{Proposition 3} \\
For a system of arbitrary number of interacting scalar fields 
minimally coupled to gravity and described by nontrivial Lagrange 
equations of motion, the full stress tensor cannot vanish for all 
solutions if it is not identically zero. \\

Proof.\\
For minimally coupled scalars any Lagrangian is of the form 
$L(\phi_a, \phi_{a,\mu}, g_{\mu\nu})$ and is free of the  metric 
connection and this results in the stress tensor free of the 
second order covariant derivatives, i.e. $T_{\mu\nu}=
T_{\mu\nu}(\phi_a, \phi_{a,\mu}, g_{\mu\nu})$. This means that on 
a Cauchy surface the stress $T_{\mu\nu}$ is determined solely by 
the initial data and thus can be given (since it does not vanish 
identically) any prescribed value (apart from some restrictions) 
independently of the field equations. By continuity, the stress 
tensor is also different from zero in some neighbourhood of the 
Cauchy surface. \\
In the case of nonminimal coupling the proposition does not apply 
directly. However, all known cases comprise the scalar field of 
scalar--tensor gravity theories (all generalizations of 
Brans--Dicke theory), the conformally invariant scalar field 
(which mathematically is merely a special case of scalar--tensor 
gravity) and a scalar field arising in restricted metric nonlinear 
gravity theories (Lagrangian being a smooth function of the 
curvature scalar) via a suitable Legendre transformation. The 
nonminimal coupling takes the form $f(\phi)R$, the Lagrangian may 
or may not contain a kinetic term for $\phi$. Then one introduces 
a new spacetime metric $\tilde{g}_{\mu\nu}$ by means of a 
Legendre map (actually this map is a conformal map of the 
original metric $g_{\mu\nu}$) and suitably redefines the 
scalar, $\tilde{\phi}=\tilde{\phi}(\phi)$. The mapping is commonly 
denoted as a transition from Jordan frame, i.e.~the system 
$(g_{\mu\nu},\phi)$, to Einstein frame consisting of 
$\tilde{g}_{\mu\nu}$ and $\tilde{\phi}$. In Einstein frame 
$\tilde{\phi}$ is minimally coupled to $\tilde{g}_{\mu\nu}$ (which 
should be regarded as the physical spacetime metric) \cite{MS2, 
ABJT} and Proposition 3 works. \\

\textbf{Proposition 4} \\
For a single vector field $A_{\mu}$ in Minkowski spacetime there 
exist solutions in the space of all solutions vanishing 
sufficiently quickly at spatial infinity for which $T_{\mu\nu}
(A)\ne 0$. \\

Proof. \\
A first order Lagrangian $L(A_{\mu}, A_{\mu,\nu})$ in Cartesian 
coordinates generates $T_{\mu\nu}(A_{\alpha}, A_{\alpha,\beta},
A_{\alpha, \beta\gamma})$ explicitly depending on second 
derivatives except the case where $L$ depends on $A_{\mu,\nu}$ 
only via $F_{\mu\nu}$. Let $S$ be a Cauchy surface with a unit 
timelike normal vector $n^{\nu}$, $n^{\nu}n_{\nu}=-1$. The initial 
data for $A_{\mu}$ form a set $C_A\equiv \{A_{\mu}, n^{\nu}
A_{\mu,\nu}\}$ of functions given on $S$. On the boundary 
2--sphere $\partial S$ at spatial infinity the initial data 
vanish.\\
One integrates the Belinfante--Rosenfeld identity (eq. (A.12) in 
Appendix), expressed in Cartesian coordinates, over $S$, 
\begin{eqnarray}\label{n47}
0\equiv \int_{S} (T^{\mu\nu}-t^{\mu\nu} +A^{\mu}E^{\nu}+
\partial_{\alpha}K^{\mu\nu\alpha})n_{\nu}\, \ud S.
\end{eqnarray}
On $S$ the tensor $K^{\mu\nu\alpha}$ being a linear combination of 
the classical spin tensor $S^{\mu\nu\alpha}$, eqs. (A.8--A.9), is 
determined by the initial data, $K^{\mu\nu\alpha}=
K^{\mu\nu\alpha}(C_A)$. Then applying the antisymmetry 
$K^{\mu\nu\alpha}=-K^{\mu\alpha\nu}$ and the Gauss' formula one 
can replace the integral of $n_{\nu}\partial_{\alpha}
K^{\mu\nu\alpha}$ over $S$ by the integral of $K^{\mu\nu\alpha}$ 
over $\partial S$ and the latter is zero. Let $A_{\mu}$ be a 
solution of $E^{\nu}=0$ corresponding to the initial data $C_A$. 
Eq. (47) reduces to 
\begin{eqnarray}\label{n48}
\int_{S} (T^{\mu\nu}-t^{\mu\nu})n_{\nu}\, \ud S=0.
\end{eqnarray}
The canonical energy--momentum tensor $t^{\mu\nu}$, eq. (A.7), is 
also determined on $S$ by the initial data, whereas the stress 
tensor is determined by the solution $A_{\mu}$. Since $t^{\mu\nu}$ 
is not zero in general, one can choose such initial data $C_A$ 
that the integral of $t^{\mu\nu}(C_A)$ does not vanish. Then the 
conserved total 4--momentum of the field for the solution is 
different from zero,
\begin{eqnarray}\label{n49}
P^{\mu}=\int_{S} T^{\mu\nu}(A)n_{\nu}\, \ud S= \int_{S}
t^{\mu\nu}(C_A) n_{\nu}\, \ud S\ne 0.
\end{eqnarray}
The Proposition is proved. \\

In this form the proof works only in Minkowski spacetime. Let 
$A_{0\mu}$ be a solution in flat spacetime for which eq. (49) 
holds. Making a small perturbation $A_{\mu}=A_{0\mu}+
\epsilon_{\mu}$ and $g_{\mu\nu}=\eta_{\mu\nu}+h_{\mu\nu}$ (where 
$h_{\mu\nu}$ may be a solution to Einstein field equations with 
some matter source, not necessarily equal to $A_{\mu}$) one finds 
that $T_{\mu\nu}$ is altered by a small quantity $T_{\mu\nu}^{L}$ 
which is linear in $\epsilon_{\mu}$ and $h_{\mu\nu}$. One then 
concludes that for spacetimes which are sufficiently close to flat 
space and for solutions being a small perturbation of $A_{0\mu}$, 
the stress tensor is nonzero too. \\

Furthermore, the proof of the Proposition may be generalized to 
at least one class of curved spacetimes. Let a spacetime metric 
admit a covariantly constant vector field $k_{\mu}$, $k_{\mu;
\nu}=0$. For example, if the spacetime is empty, $R_{\mu\nu}=0$ 
(and $d=4$), then $k_{\mu}$ is null and the spacetime represents 
the plane--fronted gravitational wave. In such a spacetime one 
considers a Cauchy surface extending to the spatial infinity. As 
previously, the initial data for $A_{\mu}$ are $C_A= \{A_{\mu},
n^{\nu}A_{\mu;\nu}\}$ with $n^{\nu}$ the unit normal vector to $S$; 
the data vanish on the boundary $\partial S$ at spatial infinity. 
The tensors $K^{\mu\nu\alpha}$ and $t_{\mu\nu}$ are determined 
on $S$ by the initial data whereas $T_{\mu\nu}$ can be 
evaluated only for the solution corresponding to $C_A$. \\
One multiplies the identity (A.12) by $k_{\mu}$ and integrates it 
over $S$ for a solution. The integral of the divergence term, 
$k_{\mu}n_{\nu}\nabla_{\alpha}K^{\mu\nu\alpha}=n_{\nu}
\nabla_{\alpha}(k_{\mu}K^{\mu[\nu\alpha]})$, is equal to the 
integral over $\partial S$ and hence is zero. One then arrives at 
an expression analogous to (48),
\begin{eqnarray}\label{n50}
\int_{S} k_{\mu}(T^{\mu\nu}-t^{\mu\nu})n_{\nu}\, \ud S=0.
\end{eqnarray}
Once again, choosing such initial data that the integral of 
$k_{\mu}t^{\mu\nu}n_{\nu}$ is nonzero, one gets $T^{\mu\nu}(A)
\ne 0$. Unfortunately, the class of spacetimes admitting a 
covariantly constant Killing field is rather narrow. \\

Finally we remark that Propositions 3 and 4 cannot be further 
strengthened, i.e.~one cannot 
exclude the situation that the stress tensor vanishes in the 
whole spacetime for some particular solutions of the Lagrange 
equations of motion. In fact, at least one counterexample is 
known: a nonlinear massive spin--two field generated by a higher 
derivative gravity theory. For this field $T_{\mu\nu}=0$ in 
the spacetime 
of a plane--fronted gravitational wave \cite{MS3}. 

\section{The canonical energy--momentum tensor in a \\ curved 
spacetime}
In many theoretical investigations in classical and quantum field 
theory in Minkowski spacetime one employs the canonical 
energy--momentum tensor due to its conceptual simplicity and 
"naturalness". Yet the variational stress tensor is constructed 
in a way which is quite artificial in flat spacetime and one must 
appeal to arguments from outside the Lagrange formalism to show 
that the latter rather than the former is the physical 
energy--momentum tensor \cite{We,HE}. \\
The canonical energy--momentum tensor for a field $\psi_A$ is  
\begin{eqnarray}\label{n51}
t_{\mu}{}^{\nu}(\psi)\equiv \delta_{\mu}^{\nu}L -
\psi_{A;\mu}\frac{\partial L}{\partial \psi_{A;\nu}}.
\end{eqnarray} 
Taking divergence of the tensor and employing Ricci identity 
(the formula after eq. (A.15)) and eq. (A.14) one finds that 
\begin{eqnarray}\label{n52}
\nabla_{\nu}t_{\mu}{}^{\nu}(\psi)=E^A \psi_{A;\mu}+
\frac{\partial L}{\partial \psi_{A;\nu}}
Z_{A}{}^{\beta}{}_{\alpha}R^{\alpha}{}_{\beta\nu\mu}.
\end{eqnarray} 
This means that even if the field equations hold, $E^A=0$, the 
canonical tensor is not conserved in a curved spacetime. This 
is the case of the electromagnetic (Maxwell) field, 
\begin{eqnarray}\label{n53}
\nabla_{\nu}t_{\mu}{}^{\nu}(A)=\frac{1}{4\pi}A_{\sigma}
F^{\alpha\beta}R^{\sigma}{}_{\alpha\beta\mu}.
\end{eqnarray}
The above and other bizarre properties of the canonical tensor 
are due to the fact that the tensor, as defined in eq. (51), does 
not fit well to the variational formalism of field theory. In fact, 
any full divergence term in the field Lagrangian does not 
contribute to $E^A$ nor to $T_{\mu\nu}(\psi)$, thus whether one 
discards such terms or not whilst evaluating these quantities 
does not affect the final outcome. Yet divergence terms in $L$ 
do contribute to the canonical tensor. For example, let for a 
vector field $L=\nabla_{\alpha}(f(A^{\mu}A_{\mu})A^{\alpha})$, 
then $t^{\mu\nu}(A)=\nabla_{\alpha}(fg^{\mu\nu}A^{\alpha}-
fg^{\mu\alpha}A^{\nu})$. \\
This example illustrates a generic feature of $t^{\mu\nu}$ for 
vector fields: if the Lagrangian is a total divergence, then 
$T_{\mu\nu}(A)\equiv 0 \equiv E^{\mu}$ and the 
Belinfante--Rosenfeld identity (A.12) yields 
\begin{eqnarray}\label{n54}
t^{\mu\nu}=\frac{1}{2}\nabla_{\alpha}(S^{\mu\nu\alpha}
+S^{\alpha\mu\nu}+S^{\alpha\nu\mu})
\end{eqnarray}
with a nonvanishing spin tensor. This vector field has no 
physical propagation and carries no energy, yet $t^{\mu\nu}$ 
apparently attributes to it some nontrivial feature. 
This fact convincingly shows that the canonical 
energy--momentum tensor is not a physical quantity. 

\section*{Appendix}

\renewcommand{\theequation}{\Alph{section}.\arabic{equation}}
\setcounter{section}{1}
\setcounter{equation}{0}

For the reader's convenience we rederive here the 
Belinfante--Rosenfeld identity for any vector field generalized 
to a curved 
spacetime since it cannot be easily found in the literature in 
the form suitable for the present paper. The identity relates the 
variational and the canonical energy--momentum tensors with the 
classical spin tensor and the Lagrange equations. Our approach is 
closest to that in \cite{KBLB, JS, BFF}, for a different approach 
see \cite{KT, Kop, HMMN}. (The Belinfante--Rosenfeld identity 
for the vacuum electromagnetic field, i.e. Maxwell's equations 
hold, in a curved spacetime is implicitly given in \cite{PK}.) 
We also recall the derivation of the 
Noether identity for any matter tensor field arising from the 
diffeomorphism invariance of the field action in the case of any 
metric theory of gravity. For an arbitrary vector matter field 
we show the explicit relationship between the generalized 
Belinfante--Rosenfeld identity and the Noether identity. \\

Let a classical vector matter field $A_{\mu}$ on a spacetime 
$(M, g_{\mu\nu})$ be described by a scalar Lagrangian $L$. The 
Lagrangian may be written either in the explicitly covariant way 
as a function of tensors or in the noncovariant form as a function 
of tensors and their partial derivatives, 
\begin{eqnarray}\label{nA.1}
L(A_{\mu}, A_{\mu;\nu}, g^{\alpha\beta})= L(A_{\mu}, A_{\mu,\nu}-
\Gamma^{\sigma}_{\nu\mu}A_{\sigma}, g^{\alpha, \beta})
\equiv L'(A_{\mu}, A_{\mu,\nu}, g^{\alpha\beta}, g^{\alpha\beta}
{}_{,\nu}).
\end{eqnarray}
The Lie derivative of the scalar density (of the weight $+1$) 
$\sqg L$ with respect to an arbitrary vector field 
$\xi^{\mu}$  may be evaluated either with the aid of the formula 
\begin{eqnarray}\label{nA.2}
{\cal L}_{\xi}(\sqg L)=\sqg \nabla_{\mu}(L\xi^{\mu})=\sqg\left[L
\xi^{\mu}{}_{;\mu}+ \xi^{\mu}\left(\frac{\partial L}
{\partial A_{\alpha}}A_{\alpha;\mu} +\frac{\partial L}
{\partial A_{\alpha;\beta}}A_{\alpha;\beta\mu}\right)\right],
\end{eqnarray}
or by Lie differentiating it as a composite function,
\begin{eqnarray}\label{nA.3}
{\cal L}_{\xi}(\sqg L) & = & {\cal L}_{\xi}(\sqg L')=
\frac{\partial(\sqg L')}{\partial A_{\mu}}{\cal L}_{\xi}A_{\mu}+
\frac{\partial(\sqg L')}{\partial A_{\mu,\nu}}{\cal L}_{\xi}
(\partial_{\nu}A_{\mu})
\nonumber\\
& & {}+\frac{\partial(\sqg L')}{\partial g^{\alpha\beta}}
{\cal L}_{\xi}g^{\alpha\beta}+ \frac{\partial(\sqg L')}
{\partial g^{\alpha\beta}{}_{,\nu}}
{\cal L}_{\xi}g^{\alpha\beta}{}_{,\nu}
\nonumber\\
&= & \left[\frac{\partial(\sqg L')}{\partial A_{\mu}}-
\partial_{\nu}\left(\frac{\partial(\sqg L')}{\partial A_{\mu,\nu}}
\right)\right]{\cal L}_{\xi}A_{\mu}
+\left[\frac{\partial(\sqg L')}{\partial g^{\alpha\beta}}
\right.
\nonumber\\
& & \left.{} -
\partial_{\nu}\left(\frac{\partial(\sqg L')}
{\partial g^{\alpha\beta}{}_{,\nu}}\right)\right]{\cal L}_{\xi}
g^{\alpha\beta}+
\partial_{\nu}\left[\frac{\partial(\sqg L')}
{\partial A_{\mu,\nu}}{\cal L}_{\xi}A_{\mu}+
\frac{\partial(\sqg L')}{\partial g^{\alpha\beta}{}_{,\nu}}
{\cal L}_{\xi}g^{\alpha\beta}\right].
\nonumber\\ & &
\end{eqnarray}
One uses the noncovariant form $L'$ in order to arrive at the 
variational stress tensor and in the other differentiations 
one replaces $L'$ by $L$. 
By applying ${\cal L}_{\xi}A_{\mu}=\xi^{\nu}A_{\mu;\nu}+
A_{\nu}\xi^{\nu}{}_{;\mu}$ and ${\cal L}_{\xi}g^{\alpha\beta}=
-2\xi^{(\alpha;\beta)}$ one recasts (A.3) to a form explicitly 
exhibiting the dependence of the Lie derivative on $\xi^{\mu}$ 
and its first and second order derivatives. First, denoting 
\begin{displaymath}
d^{\alpha\beta}\equiv \frac{\partial L}{\partial A_{\alpha;
\beta}} \qquad \textrm{and} \qquad 
E^{\mu}[L(A)]\equiv \frac{\partial L}{\partial A_{\mu}}-
\nabla_{\alpha}\left(\frac{\partial L}{\partial A_{\mu;\alpha}}
\right),
\end{displaymath}
the Euler operator, one finds
\begin{displaymath}
\frac{\partial(\sqg L')}{\partial A_{\mu}}=\sqg
\left(\frac{\partial L}{\partial A_{\mu}}+
\frac{\partial L}{\partial A_{\alpha;\beta}}
\frac{\partial A_{\alpha;\beta}}{\partial A_{\mu}}\right)=
\sqg\left(\frac{\partial L}{\partial A_{\mu}}-
d^{\alpha\beta}\Gamma^{\mu}_{\alpha\beta}\right)
\end{displaymath}
\begin{displaymath}
\textrm{and} \qquad \partial_{\nu}\left(\frac{\partial(\sqg L')}
{\partial A_{\mu,\nu}}\right)=
\sqg(d^{\mu\nu}{}_{;\nu}-
d^{\alpha\beta}\Gamma^{\mu}_{\alpha\beta}).
\end{displaymath}
Then the first square bracket in (A.3) equals $\sqg E^{\mu}$. The 
second square bracket is just $-\frac{1}{2}\sqg T_{\alpha\beta}
(A)$. The first term in the third square bracket is also simple, 
\begin{displaymath}
\frac{\partial(\sqg L')}{\partial A_{\mu,\nu}}=
\sqg \frac{\partial L}{\partial A_{\mu;\nu}},
\end{displaymath}
whereas the second term in this bracket is a complicated tensor 
density. After several manipulations with the last bracket and 
upon equating (A.2) to (A.3) one gets a scalar identity of the 
form 
\begin{eqnarray}\label{nA.4}
\xi^{\mu}P_{\mu}+\xi_{\mu;\nu}Q^{\mu\nu}+
\xi_{\alpha;(\mu\nu)}R^{\alpha(\mu\nu)}\equiv 0.
\end{eqnarray}
The identity holds for any vector fields $A_{\mu}$ and 
$\xi^{\mu}$. At any spacetime point the tensors $\xi_{\mu}$, 
$\xi_{\mu;\nu}$ and $\xi_{\alpha;(\mu\nu)}$ are independent 
quantities and therefore their coefficients must identically 
vanish. Thus the identity splits into a cascade of $4+16+40$ 
identities. The identities $R^{\alpha(\mu\nu)}\equiv 0$ are 
purely algebraic functions of of $A_{\mu}$, $g_{\mu\nu}$ and 
$d^{\mu\nu}$ and are completely trivial in the sense that they 
hold for any values of these 3 tensors. The identities $P_{\mu}
\equiv 0$ are also algebraic functions of these tensors and 
$R_{\alpha\beta\mu\nu}$ (arising from $\xi_{\alpha;[\mu\nu]}$) 
and are trivially satisfied provided $d^{\mu\nu}$ equals 
$\partial L/\partial A_{\mu;\nu}$. Only $Q^{\mu\nu}\equiv 0$ 
are nontrivial and read 
\begin{eqnarray}\label{nA.5}
Q^{\mu\nu}=T^{\mu\nu}+A^{\mu}E^{\nu}- Lg^{\mu\nu}+
A_{\alpha}{}^{;\mu}d^{\alpha\nu}+
\nabla_{\alpha}K^{\mu\nu\alpha}\equiv 0,
\end{eqnarray}
where
\begin{eqnarray}\label{nA.6}
K^{\mu\nu\alpha}\equiv A^{\mu}N^{\nu\alpha} -S^{\mu\alpha}
A^{\nu} +S^{\mu\nu}A^{\alpha}
\end{eqnarray}
and $S^{\mu\nu}\equiv d^{(\mu\nu)}$ and $N^{\mu\nu}\equiv 
d^{[\mu\nu]}$. One introduces the Pauli's canonical 
energy--momentum tensor 
\begin{eqnarray}\label{nA.7}
t_{\mu}{}^{\nu}(A)\equiv \delta^{\nu}_{\mu}L -A_{\alpha;\mu}
\frac{\partial L}{\partial A_{\alpha;\nu}} 
\end{eqnarray}
and splits $K^{\mu\nu\alpha}$ into its symmetric and 
antisymmetric parts w.r.t.~to the first two indices,
\begin{eqnarray}\label{nA.8}
K^{[\mu\nu]\alpha}=\frac{1}{2}(A^{\mu}d^{\nu\alpha}-
A^{\nu}d^{\mu\alpha})\equiv \frac{1}{2}S^{\mu\nu\alpha},
\end{eqnarray}
\begin{eqnarray}\label{nA.9}
K^{(\mu\nu)\alpha}=
A^{\alpha}d^{(\mu\nu)}-d^{\alpha(\mu}A^{\nu)}=
S^{\alpha(\mu\nu)}.
\end{eqnarray}
The tensor $S^{\mu\nu\alpha}$ may be interpreted as the density 
of the classical spin (helicity) of the vector field. In fact, 
for the electromagnetic field in vacuum in Minkowski spacetime 
the spin density tensor is defined as 
\begin{eqnarray}\label{nA.10}
S^{\mu\nu\alpha}_{\textrm{EM}}=x^{\mu}(T^{\nu\alpha}-
t^{\nu\alpha}) -x^{\nu}(T^{\mu\alpha}-t^{\mu\alpha}).
\end{eqnarray}
For $L_{\textrm{EM}}=-1/(16\pi)F_{\alpha\beta}F^{\alpha\beta}$ 
and assuming that Maxwell equations hold, $F^{\alpha\beta}{}_
{,\beta}=0$, and the field vanishes sufficiently quickly at 
spatial infinity, one can remove a full divergence term from the 
spin density tensor since it does not contribute to the total 
spin of the electromagnetic field. As a result, 
\begin{eqnarray}\label{nA.11}
S^{\mu\nu\alpha}_{\textrm{EM}}=\frac{1}{4\pi}
(A^{\mu}F^{\nu\alpha}-A^{\nu}F^{\mu\alpha})=
A^{\mu}d^{\nu\alpha}-A^{\nu}d^{\mu\alpha},
\end{eqnarray}
\cite{BoS,KBLB}. The tensor $K^{\mu\nu\alpha}$ is antisymmetric, 
$K^{\mu\nu\alpha}=-K^{\mu\alpha\nu}$. Finally (A.5) takes on the 
form
\begin{eqnarray}\label{nA.12}
Q^{\mu\nu}=T^{\mu\nu}- t^{\mu\nu}+A^{\mu}E^{\nu}+\frac{1}{2}
\nabla_{\alpha}(S^{\mu\nu\alpha}+S^{\alpha\mu\nu}+
S^{\alpha\nu\mu})\equiv 0.
\end{eqnarray}
These are the Belinfante--Rosenfeld identities \cite{Be,Ro} for 
any vector field, generalized to a curved spacetime. Notice that 
the identities arise due to the fact that the field Lagrangian is 
a scalar, hence the action is invariant under infinitesimal 
coordinate transformations (with appropriate boundary 
conditions). \\

The proof of Propositions 1 and 2 in sect. 2 is based on the 
famous second Noether theorem (also named "the Noether 
identities", "the strong conservation laws" or "Bianchi 
identities for matter") \cite{T,AD,FF}; our approach is based on 
\cite{FF,MS1}. Let $\psi_A$ denote an arbitrary tensor matter 
field (or a set of tensor fields) with a collective index $A$, 
described by a scalar Lagrangian $L(\psi)=L(\psi_A, \psi_{A;\mu}, 
g_{\mu\nu}, R_{\alpha\beta\mu\nu})$, i.e.~one admits a possible 
non--minimal coupling to the curvature. The action integral for 
for Einstein gravity (actually one may envisage any metric theory 
of gravity since the derivation goes then without alteration) and 
the matter field is (we set $c=8\pi G=1$)
\begin{eqnarray}\label{nA.13}
S=\int_{\Omega} (\frac{1}{2}R+L)\sqg\, \ud^4 x. 
\end{eqnarray}
The Euler operator for the equations of motion for $\psi_A$ is 
\begin{eqnarray}\label{nA.14}
E^{A}[\psi,g]\equiv \frac{\partial L}{\partial \psi_{A}}-
\nabla_{\mu}\left(\frac{\partial L}{\partial \psi_{A;\mu}}
\right).
\end{eqnarray}
The action (A.13) is diffeomorphism invariant and in particular 
it remains unchanged by an infinitesimal point transformation 
$x'^{\mu}=x^{\mu}+\xi^{\mu}(x)$ with arbitrary $\xi^{\mu}$ which 
vanishes on the boundary of the domain $\Omega$. The integration 
domain is thus mapped onto itself whereas the variations of the 
fields are given by Lie derivatives, $\delta\psi_{A}= -{\cal L}_
{\xi}\psi_{A}$ and $\delta g^{\mu\nu}=-{\cal L}_{\xi}
g^{\mu\nu}$. These variations can be expressed in terms of 
coefficients $Z_{A}{}^{\beta}{}_{\alpha}(\psi)$, being linear 
functions of $\psi_{A}$, 
\begin{eqnarray}\label{nA.15}
{\cal L}_{\xi}\psi_{A}=\xi^{\alpha}\psi_{A;\alpha}+
Z_{A}{}^{\beta}{}_{\alpha}\xi^{\alpha}{}_{;\beta},
\end{eqnarray}
which also appear in the covariant derivative,
\begin{displaymath}
\nabla_{\mu}\psi_{A}=\partial_{\mu}\psi_{A}-
Z_{A}{}^{\beta}{}_{\alpha}(\psi)\Gamma^{\alpha}_{\mu\beta}
\end{displaymath}
and in Ricci identity,
\begin{displaymath}
\psi_{A;\mu\nu}-\psi_{A;\nu\mu}=
Z_{A}{}^{\beta}{}_{\alpha}(\psi)R^{\alpha}{}_{\beta\mu\nu}.
\end{displaymath}
The invariance of the action implies 
\begin{eqnarray}\label{nA.16}
0=\delta S= \int_{\Omega} [\frac{1}{2}(G_{\mu\nu}-T_{\mu\nu})
\delta g^{\mu\nu} +E^A \delta \psi_A]\sqg\, \ud^4 x 
\end{eqnarray}
plus a surface integral which vanishes due to the boundary 
conditions. Applying (A.15), dropping again a total divergence 
and making use of the ordinary Bianchi identity $G_{\mu}^{\nu}
{}_{;\nu}\equiv 0$ one finds that the integrand is of the form 
$\xi^{\mu}B_{\mu}(\psi)$. Vanishing of the integral and 
arbitrariness of $\xi^{\mu}$ in the interior of $\Omega$ entail 
$B_{\mu}\equiv 0$ or 
\begin{eqnarray}\label{nA.17}
E^A\psi_{A;\alpha}\equiv \nabla_{\beta}(T_{\alpha}{}^{\beta}+
E^A Z_{A}{}^{\beta}{}_{\alpha}).
\end{eqnarray}
This is the Noether identity for any classical tensor matter 
field, valid for any dimension $d\geq 3$. Notice that the second 
Noether theorem derived in refs. \cite{BCJ,JS} does not include 
the variational (stress) energy--momentum tensor $T_{\mu\nu}
(\psi)$. The tensor is absent there because it is (implicitly) 
assumed in these papers that the matter action is invariant under 
a symmetry transformation of the matter field (a gauge 
transformation) alone and the metric is left unaltered. \\

In the case of a vector field $A_{\mu}$ the coefficients are 
$Z_{\mu}{}^{\beta}{}_{\alpha}=\delta^{\beta}_{\mu}A_{\alpha}$ 
and the four Noether identities 
\begin{eqnarray}\label{nA.18}
E^{\mu}F_{\alpha\mu}\equiv T_{\alpha\beta}{}^{;\beta}+
E^{\mu}{}_{;\mu}A_{\alpha}, 
\end{eqnarray}
where $F_{\alpha\beta}\equiv A_{\beta;\alpha}-A_{\alpha;\beta}$, 
can be derived from the Belinfante--Rosenfeld identities (A.12). 
In fact, a direct and quite uphill calculation proves that 
\begin{eqnarray}\label{nA.19}
Q_{\alpha\mu}{}^{;\mu}= T_{\alpha\mu}{}^{;\mu} -
F_{\alpha\mu}E^{\mu}+A_{\alpha}E^{\mu}{}_{;\mu}\equiv 0. 
\end{eqnarray}
Analogous relationships exist for tensor fields. 

\section*{Acknowledgments}
I am grateful to Andrzej Staruszkiewicz for stimulating 
comments and suggestions. I have benefitted a lot from 
discussions with him. I thank Henryk Arod\'{z}, Wojciech 
Kopczy\'{n}ski, Jacek 
Jezierski and Jerzy Kijowski for their comments on this work.  
This work was partially supported 
by KBN grant no. 2P03D01417.

\end{document}